# Recovering the Intrinsic Metallicity Distribution of Elliptical Galaxies

L. Ciotti[1,2], M. Stiavelli[1,3] and A. Braccesi[4]
[1] *Dept. of Theoretical Physics, Oxford University, UK*
[2] *Osservatorio Astronomico di Bologna, I40100 Bologna, Italy*
[3] *Scuola Normale Superiore, I56126 Pisa, Italy*
[4] *Dipartimento di Astronomia, Università di Bologna, I40100 Bologna, Italy*

22 May 1995

**ABSTRACT**
We address the problem of deriving, from the observed projected metallicity gradients, the intrinsic metallicity distribution of elliptical galaxies as a function of their integrals of motion. The method is illustrated by an application to anisotropic spherical Hernquist models. We also compare the derived metallicity distribution with those expected from two very simple models of galaxy formation and find that the more dissipative scheme agrees better with the typical metallicity distribution of ellipticals.

## 1 INTRODUCTION

The aim of this paper is to give a simple recipe to derive the metallicity distribution of galaxies as a function of their integrals of the motion. Elliptical galaxies are known to possess metallicity gradients (Peletier 1989, Gorgas, Efstathiou and Salamanca 1990, Davies, Sadler and Peletier 1993, Carollo, Danziger and Buson 1993), established mainly by measuring the $Mg_2$ line strength, which is one of the most prominent features in the optical spectra of early type galaxies and K and M type giants. The observed metallicities, as determined by this line frequently show variations of a factor 2 from the center to the effective radius. The observed gradients are the result of *projection* and *orbital mixing*. The former is easy to model and was in fact shown by Peletier (1989) to lead to only minor changes in the observed metallicity gradients. The latter is more complex and has not been studied in detail. Orbital mixing arises because stars at a given radius may have their apocenters spanning a wide range of radii. Thus, the mean metallicity at any given point inside the galaxy is the result of the metallicity distribution in phase space weighted by the galaxian distribution function.

In order to investigate these effects, we develop a technique to derive the metallicity distribution in phase space from the observed metallicity gradients of for spherical galaxies. The technique is a simple extension of Eddington's inversion to derive the distribution function from the galaxy density profile. The method is described in Sect. 2 and 3, and applied to anisotropic spherical Hernquist models in Sect. 4. In Sect. 5 we compare the derived phase space metallicity distribution to the one predicted by simple models of galaxy formation (Sect. 5), finding the simple prediction of dissipative models in better agreement with observations.

## 2 FROM $Z_P$ TO $Z_D$

For a spherical galaxy the luminosity density $\nu(r)$ is obtained by deprojecting its surface brightness profile $I(R)$:

$$\nu(r) = -\frac{1}{\pi}\int_r^\infty \frac{dI(R)}{dR}\frac{dR}{\sqrt{R^2-r^2}}, \qquad (1)$$

where $r$ is the spatial radius, and $R$ the projected one (see, e.g., Binney and Tremaine 1987, hereafter BT). The observed (projected) metallicity $Z_P(R)$ is the result of a projection along the line-of-sight of the luminosity-weighted spatial *mean* metallicity $Z_D(r)$:

$$I(R)Z_P(R) = 2\int_R^\infty \frac{\nu(r)Z_D(r)r\,dr}{\sqrt{r^2-R^2}}, \qquad (2)$$

which can be inverted to yield:

$$\nu(r)Z_D(r) = -\frac{1}{\pi}\int_r^\infty \frac{dI(R)Z_P(R)}{dR}\frac{dR}{\sqrt{R^2-r^2}}. \qquad (3)$$

Note that $Z_D(r)$ as derived from the ratio of Eqs. (3) and (1), *does not depend* on the unknown galaxy mass-to-light ratio $\Upsilon(r) = \rho(r)/\nu(r)$.

## 3 RECOVERING Z(Q)

Assuming that the previous galaxy is described by a distribution function $f = f(Q)$, the relation between its density $\rho$ and $f$ is

$$\rho_Q(r) \equiv \rho(r)\left(1 + \frac{r^2}{r_a^2}\right) = 4\pi \int_0^\Psi f(Q)\sqrt{2(\Psi - Q)}dQ, \quad (4)$$

where $Q = \mathcal{E} - L^2/2r_a^2$, $\mathcal{E} = \Psi(r) - v^2/2$ is the stellar binding energy per unit mass, $\Psi(r)$ is the gravitational potential, $L^2$ is the square modulus of the angular momentum, and $r_a$ is the so-called *anisotropy radius* (Osipkov, 1979; Merritt, 1985; see also BT for a general account). For $r_a = \infty$ the distribution function depends on the energy only, and the velocity dispersion tensor is globally isotropic. For $r_a = 0$ all the stellar orbits are radial. Equation (4) can be inverted, namely (see BT):

$$f(Q) = \frac{1}{\sqrt{8}\pi^2}\frac{d}{dQ}\int_0^Q \frac{d\rho_Q}{d\Psi}\frac{d\Psi}{\sqrt{Q-\Psi}}. \quad (5)$$

Therefore, once $\Upsilon(r)$ is known, one can *formally* derive, for any fixed anisotropy radius, an equilibrium distribution function $f(Q)$ for a spherically symmetric system with a given $I(R)$.

The quantity $Z_D(r)$ is the *mean* metallicity of all stars observed at the radius $r$; these stars can be characterized by very different $Q$ values, i.e. their orbits lie in different regions of the phase space. In order to derive the *intrinsic* metallicity distribution in phase space, the function $Z(Q)$, one proceeds in a similar way that in Eqs. (1)-(3). The metallicity dependence of the mass-to-light ratio of a stellar population, $\Upsilon = \Upsilon(Z)$, gives the two equations:

$$\nu_Q(r) = 4\pi \int_0^\Psi f(Q)\Upsilon^{-1}(Z)\sqrt{2(\Psi - Q)}dQ, \quad (6)$$

and

$$\nu_Q(r)Z_D(r) = 4\pi \int_0^\Psi f(Q)\Upsilon^{-1}(Z)Z(Q)\sqrt{2(\Psi - Q)}dQ. \quad (7)$$

Their functional inversion is the same as in Eq. (5):

$$f(Q)\Upsilon^{-1}(Z) = \frac{1}{\sqrt{8}\pi^2}\frac{d}{dQ}\int_0^Q \frac{d\nu_Q}{d\Psi}\frac{d\Psi}{\sqrt{Q-\Psi}}. \quad (8)$$

and

$$f(Q)\Upsilon^{-1}(Z)Z(Q) = \frac{1}{\sqrt{8}\pi^2}\frac{d}{dQ}\int_0^Q \frac{d\nu_Q Z_D}{d\Psi}\frac{d\Psi}{\sqrt{Q-\Psi}}. \quad (9)$$

The functions needed for the inversion are provided by Eq. (1) and Eq. (3), and we can derive separately the product $f(Q)\Upsilon^{-1}(Z)$ and the intrinsic metallicity $Z(Q)$, which *does not depend on the knowledge of* $\Upsilon$. Moreover, if we can determine $\Upsilon$ from synthetic stellar population, e.g., for a 12 Gyrs population:

$$\log_{10}\Upsilon_V \simeq 1.178 + 0.1927\log_{10}Z \quad (10)$$

(Peletier, 1989), then from Eq. (8) we can obtain also $f$, and from that the projected velocity dispersion $\sigma_P(R)$. A difficulty may arise if the derived $Z(Q)$ and $f(Q)$ are negative for some permitted value of $Q$. Indeed, it is well known that the previous inversion procedure does not guarantee the positivity of the recovered function.

## 4  A TOY MODEL

We have applied the method to the case of Hernquist models (Hernquist, 1990), that approximate the de Vaucouleurs

**Figure 1.** Left panel: projected (solid line) and deprojected (dotted line) metallicity profiles as a function of radius. Right panel: the derived intrinsic metallicity profile is shown, for the isotropic model (solid line), and for anisotropic model with $r_a = R_e$. The two short-dashed curves refer to the dissipationless galaxy formation models described in Sect. 5, and the straight, long-dashed line refers to a simple dissipative formation model.

$R^{1/4}$ law, and for which many properties can be evaluated analytically. The model investigated is characterized by a total mass $10^{11} M_\odot$, an effective radius $R_e = 4$ kpc, and a projected metallicity profile given by

$$Z_P(R) = Z_\odot \left[\left(\frac{Z_P(0)}{Z_\odot}\right)^{0.41} - 0.18\log_{10}\left(1 + \frac{R}{R_Z}\right)\right]^{2.439}, (11)$$

which agrees with the observations of $Mg_2$ gradients in ellipticals calibrated to metallicity by using equation (1) of Bender, Burstein and Faber (1993) for a population 12 Gyrs old. Although steeper gradients will have a larger effect, we consider this slope to be representative of the mean gradients measured in ellipticals. Inside $R_Z = R_e/30$, $Z_P(R)$ is taken to assume the central value $Z_P(0) = 2Z_\odot$, where $Z_\odot \simeq 0.02$ is the solar metallicity. Our results are shown in Fig. 1, where in the left panel we show the projected and the deprojected metallicity profiles, $Z/Z_\odot$, as a function of radius. In the right panel we show the intrinsic metallicity of stars as a function of $Q$, for a globally isotropic model and for an anisotropic model characterized by $r_a = R_e$. Note that the phase-space metallicity distribution obtained is positive everywhere, and, for the isotropic case, nicely close to a straight line over a wide interval in $Q$. The same linear behavior is obtained also for steeper and flatter metallicity gradients (not shown in Fig. (1)). In the next section we will discuss some implications of this result.

## 5  COMPARISON WITH FORMATION MODELS

The formalism derived above also allows one to set some constraints on galaxy formation models. Rather than considering a given galaxy, here we focus on the model described above and taken to be *representative* of typical ellipticals. As a first comparison of the derived intrinsic metallicity $Z(\mathcal{E})$ (here we consider only the isotropic case, where $Q$ reduces to the binding energy $\mathcal{E}$) with formation models, we have considered two different, extreme but simple, scenarios. The first scenario is based on the assumption that galaxies collapse dissipatively and, after they have essentially settled on their final density profile, form stars. The binding energy distribution in such a system would match closely the gravitational potential. In this scenario one would expect $Z(\mathcal{E}) \propto \mathcal{E}$. This is the underlying assumption leading several authors to compare metallicities with escape velocities (see, e.g., Franx and Illingworth 1988, Davies et al. 1993, Carollo and Danziger 1994). Despite the simplicity of the argument and the lack of detailed modelling, these authors confirmed that a good correlation is indeed observed between (projected) metallicities and escape velocities averaged along the line of sight.

The second scenario is based on the assumption that galaxies collapses without dissipation after having formed their stars on timescales shorter to dynamical ones. We assume that stars form out of a top hat gas cloud of constant density $\rho_0$. Such a model has a gravitational potential given by:

$$\Psi(\xi) = \Psi(0)(1 - \frac{1}{3}\xi^2), \qquad (12)$$

with $\Psi(0)$ and $\xi$ being, respectively, the central value of the gravitational potential and the radius relative to the outer boundary of the top hat disturbance.

At each radius, star formation and enrichment proceed following the simple model illustrated in, e.g., BT, until the gas becomes unbound due to SN energy release. We assume that each radial shell evolves as a closed box model. The metallicity profile of stars in the (pre-collapse) protogalaxy is then given, in terms of the gas metallicity, by:

$$Z_\star(\xi) = p \frac{1 - [1 + Z_{gas}(\xi)/p] \exp[-Z_{gas}(\xi)/p]}{1 - \exp -[Z_{gas}(\xi)/p]}, \qquad (13)$$

where $p$ is the chemical yield. In the limit of very large $Z_{gas}/p$ Eq. (13) gives $Z_\star = p$. In the opposite limit, $Z_{gas}/p \ll 1$ one has $Z_\star = Z_{gas}/2$, which implies $p \gg 2Z_\star$. Since the element production of the supernovae is proportional to their energy release, and, at the onset of the galactic wind, the energy release per unit mass equals the local gravitational potential, one has:

$$Z_{gas}(\xi) = Z_{gas}(0) \frac{\Psi(\xi)}{\Psi(0)}. \qquad (14)$$

Note that the stars that have formed do not acquire significant velocities while star formation is active, i.e., $\mathcal{E} = \Psi(\xi)$, since the basic assumption of the dissipationless collapse scheme is that the star formation time scale is shorter than the dynamical time scale. Assuming that the evolution of gravitational potential during the wind phase can be neglected, the stellar density, after the wind has occurred, is obtained from Eq. (9-21) of BT, namely:

$$\frac{\rho_\star(\xi)}{\rho_0} = 1 - \exp\left[-\frac{Z_{gas}(0)}{p} \frac{\Psi(\xi)}{\Psi(0)}\right], \qquad (15)$$

where $Z_{gas}(0)$ is the central gas metallicity. This allows us to compute the cumulative energy distribution at the occurrence of the wind, i.e. the mass fraction of stars more bound than a given energy. By assuming that the subsequent dissipationless collapse does not alter this distribution, as actually verified in N-body simulations, one predicts the final $Z(\mathcal{E})$. This is accomplished by remapping the initial cumulative energy distribution into that of the an Hernquist sphere with the same total mass.

The predictions from these two models are shown in Fig. 1 (right panel, dashed lines) where they are compared to the intrinsic metallicity derived by our inversion procedure. The two short dashed lines represent the top hat gas cloud model prediction for the intrinsic metallicity of a galaxy with a final central star metallicity of $2Z_\odot$. The two lines correspond to the minimum value of $p$ giving the desired central stellar metallicity, $p = 0.04$, and to $p = 0.1$. All curves for $p \gg 0.04$ are similar. Note that the simple dissipative scheme (Fig. 1, long dashed line) appears to better support the typical observed gradient than the simplified dissipationless scheme.

## 6 CONCLUSIONS

We have proposed a simple inversion procedure allowing one to derive the dependence of metallicity on the integrals of motion for a spherical galaxy, under the assumption of a Osipkov-Merritt orbital anisotropy distribution. Although applied only to an analytical Hernquist model, the technique described here could be applied to real (close to spherical) galaxies for which surface brightness and metallicity profiles are available. The ideas of this paper can be generalized to oblate two-integrals models constructed following the Hunter & Qian (1993) technique. The intrinsic metallicity as a function of binding energy and angular momentum would then be obtained by inversion from a two-dimensional $Mg_2$ map.

We have also attempted a cursory discussion of the predictions of two simple galaxy formation scenarios for the intrinsic metallicity distribution as a function of the binding energy. We conclude that for typical ellipticals the more dissipative scheme is to be preferred to the dissipationless scheme. In order to derive more firm conclusions, the metallicity derived from gas dynamical simulations should be considered.

## 7 ACKNOWLEDGEMENTS

L.C. and M.S. wish to thank Giuseppe Bertin, James Binney, Reynier Peletier, Alvio Renzini, and Tjeerd van Albada for useful discussions, and the Theoretical Physics Dept., Oxford University, where this project has been partially carried out, for the warm hospitality. This work has been partially supported by SERC of UK and MURST of Italy. L.C. has been partially supported by the EEC contract No. CHRX-CT92-0033.